\documentclass{ws-procs961x669}            
\begin{document}
\title{Hyper-Fast Positive Energy Warp Drives}

\author{E. W. Lentz$^*$}

\address{Pacific Northwest National Laboratory, Richland, WA 99354 USA\\
$^*$E-mail: erik.lentz@pnnl.gov}

\begin{abstract}
Solitons in space–time capable of transporting time-like observers at superluminal speeds have long been tied to violations of the weak, strong, and dominant energy conditions of general relativity. This trend was recently broken by a new approach that identified soliton solutions capable of superluminal travel while being sourced by purely positive energy densities. This is the first example of hyper-fast solitons satisfying the weak energy condition, reopening the discussion of superluminal mechanisms rooted in conventional physics. This article summarizes the recent finding and its context in the literature. Remaining challenges to autonomous superluminal travel, such as the dominant energy condition, horizons, and the identification of a creation mechanism are also discussed.
\end{abstract}


\bodymatter

\section{Introduction}

One of the most prominent critiques of superluminal travel within Einstein's general relativity (GR) is that any geometry that facilitates such travel must be largely sourced by a form of negative energy density \cite{Alcubierre1994,Everett1996,Pfenning1997,Hiscock1997,Krasnikov1998,Olum1998,VanDenBroeck1999,Millis1999,Visser2000,Loup2001,Natario2002,Gauthier2002,Lobo2003,Lobo2004,Lobo2007,Obousy2008,Finazzi2009,White2013}. Other concerns include the immense (magnitude) energy requirements to create a soliton, the difficulty associated with constructing a soliton from a nearly flat spacetime up to the superluminal phase, where the transported central observers become surrounded by a horizon, and the equal difficulties of driving the superluminal phase back the nearly flat spacetime. 

There has been a recent uptick in interest regarding warp drives due to a set of papers made available in 2020 and 2021 claiming the construction of solutions that do not require sources with negative energy density \cite{Lentz2020,Bobrick2021,SantosPereira2021,Fell2021}, obeying the weak energy condition (WEC). These increasingly physical warp drives were the focus of a mini-session at the 16$^{th}$ Marcel Grossmann meeting that covered the history of warp drive research in academia, the recent positive energy warp drive research papers, and two articles regarding their reception \cite{Santiago2021a,Santiago2021b}. This article will concern only the paper written by the author \cite{Lentz2020}, summarizing its findings and discussing its standing in the literature. Mathematical notation will largely follow that of the original paper.

\section{Nat\'ario Class Spacetimes}

The class of relativistic spacetime metrics describing warp drive geometries in the literature are decomposed according to ``3+1'' Arnowitt-Deser-Misner (ADM) formalism~\cite{ADM}
\begin{equation}
    ds^2 = -\left(N^2-N^i N_i \right) dt^2 - 2 N_i dx^i dt + h_{ij} dx^i dx^j,
\end{equation}
where the time coordinate $t$ stratifies spacetime into space-like hypersurfaces, the space metric components $h_{ij}$ evaluated at $t$ provide the intrinsic geometry of that hypersurface, and the similarly-evaluated shift vector components $N^i$ at $t$ provide the coordinate three-velocity of the hypersurface's normal.  The time-like unit normal one-form is therefore proportional to the coordinate time element $\mathbf{n}^* = N dt$, and the unit normal vector $\mathbf{n}$ to the hypersurface has components
\begin{equation}
    n^{\nu} = \left(\frac{1}{N}, \frac{N^i}{N}  \right).
\end{equation} 
For simplicity, we will use natural units $G=c=1$. 

The majority of previous warp drive papers including Lentz 2021~\cite{Lentz2020} set the lapse function $N$ to unity and the hypersurface metric to be flat under Cartesian coordinates $h_{ij} = \delta_{ij}$. The non-flat geometry is therefore encoded in the three-component shift vector, $N_i$. The class of spacetimes described in this way have been coined as ``Nat\'ario spacetimes'' \citep{Bobrick2021,Santiago2021a,Santiago2021b}.

The projection of the Einstein equation onto the hypersurface normal gives the Hamiltonian constraint of a Nat\'ario spacetime 
\begin{equation}
    G^{\mu \nu} n_{\mu} n_{\nu} = 8 \pi T^{\mu \nu} n_{\mu} n_{\nu},
\end{equation}
with the projected stress-energy being referred to as the Eulerian energy density
\begin{equation}
    T^{\mu \nu} n_{\mu} n_{\nu} = T^{00} = E.
\end{equation}
The geometric side of this energy constraint equation can be expressed in terms of the extrinsic curvature's trace $K = K^i_i$ and its quadratic hypersurface scalar $K^i_j K^j_i$
\begin{equation}
    8 \pi E = \frac{1}{2}\left(- K^i_j K^j_i + K^2 \right).
\end{equation}
The combination of extrinsic curvatures expanded in terms of the shift vector components take the form
\begin{align}
    K^2 - K^i_j K^j_i  &= 2 \partial_x N_x \partial_y N_y + 2 \partial_x N_x \partial_z N_z + 2 \partial_z N_z \partial_y N_y \nonumber \\
    &- \frac{1}{2} \left( \partial_x N_y + \partial_y N_x  \right)^2 - \frac{1}{2} \left( \partial_x N_z + \partial_z N_x  \right)^2 - \frac{1}{2} \left( \partial_z N_y + \partial_y N_z  \right)^2. \label{Egeom}
\end{align}

The warp drive solution of Alcubierre~\cite{Alcubierre1994} set the precedent for WEC violation by requiring negative Eulerian energy throughout. Specifically, utilizing only a single component of the shift vector in the direction of motion, here taken to be along the positive z-axis, produces the renowned toroid of negative energy density about the soliton bubble of $N_z$, here displayed in Cartesian coordinates,
\begin{equation}
    E_{\text{Alc}} = \frac{-1}{32 \pi} \left( \left( \partial_x N_z  \right)^2 + \left( \partial_y N_z \right)^2 \right).
\end{equation}
The expansionless ($K = -1/2(\partial_x N_x + \partial_y N_y + \partial_z N_z) = 0$) elliptic relation of Nat\'ario 2002~\cite{Natario2002} restricted the energy form to the negative definite square of the extrinsic curvature
\begin{equation}
    E_{\text{Nat}} = \frac{-1}{16 \pi} K^i_j K^j_i.
\end{equation}
Parabolic and hyperbolic relations remained to be explored at the start of 2020.

\section{Positive Energy Warp Drives}

The soliton geometry of Lentz 2021~\cite{Lentz2020} distinguishes itself from the previous literature in that it satisfies the WEC, even when moving at superluminal speeds. The WEC states that the energy of a spacetime is nowhere negative for any time-like observer. Mathematically, this means that for any time-like vector field $X$, the projection with the stress-energy tensor $T$ must be non-negative
\begin{equation}
    X_{\mu} T^{\mu \nu} X_{\nu} \ge 0.
\end{equation}
The positive-energy soliton was identified through the construction of a set of rules sufficient to define geometries with everywhere non-negative energy. A brief presentation of the rules is given below.

Recall the expansion of the Hamiltonian constraint of Eqn.~\ref{Egeom}. Observe that the last three elements of the above expression are negative definite, while the first three are of indeterminant type. These first three terms provide opportunity for the Eulerian energy density function to be non-negative under particular configurations, so long as they are everywhere dominate over the first three terms. The next steps focus on such configurations. 

The first rule is to reduce the 3D shift vector field to a single potential function, a real-valued function $\phi$ with spatial gradient relating the shift vector components
\begin{equation}
    N_i = \partial_i \phi,
\end{equation}
satisfying a linear wave equation over the spatial coordinates
\begin{equation}
    \partial_x^2 \phi +\partial_y^2 \phi -\frac{2}{v_h^2} \partial_z^2 \phi = \rho,
\end{equation}
where $ v_h/\sqrt{2}$ is the dimensionless wave front `speed' on the hypersurface, and $\rho$ is the wave equation source function, not to be confused with mass or energy density. This step provides nearly all the structure needed to find the first example positive energy drives.

Two more simplifications are used to set sufficient rules for positive Eulerian energy. The energy functional is reduced to a two-coordinate form $(z,x)$ by restricting $\rho$ and $\phi$ to be parameterized in the ($x,y$) plane by the $l_1$ norm $s= |x| + |y|$,
\begin{equation}
    E  = \frac{1}{16 \pi} \left(2 \partial_z^2 \phi \left(\rho + \frac{2}{v_h^2} \partial_z^2 \phi \right) - 4 \left( \partial_z \partial_x \phi \right)^2 \right), \label{Edens}
\end{equation}
which can be bounded from below by
\begin{align}
    16 \pi E &\ge 2  \rho \times \partial_z^2 \phi \nonumber \\
    &= \rho \times\frac{1}{ 2 v_h} \int_{-\infty}^{\infty} dx' \partial_r \rho(r,|x'| + |y|)|_{r= z-|\Delta x|/v_h}, \label{ineq}
\end{align}
where the Green's expressions for the shift vector is used in the last expression. From the lower bound expression, the last rule is formed to ensure the Eulerian energy density is everywhere non-negative: the energy function will be non-negative for configurations such that the local source density and the $z$-component source density gradient integrated along the intersecting `past' wave trajectories are of the same sign. In other words, the two factors in Expr.~\ref{ineq} must have the same sign.

It is from this sequence of rules that the Eulerian energy can be constrained to be non-negative. Demonstrating the fullness of the WEC takes several additional steps to understand the contributions of the Eulerian momentum and stress components, and is covered in detail in the original publication \cite{Lentz2020}, but are omitted here in the interest of space. The rules invoked are not strictly necessary to positive-energy warp drives. The solution space of physical warp drives is expected to be much larger and more diverse.

\begin{figure}
\begin{center}
\includegraphics[width=\textwidth]{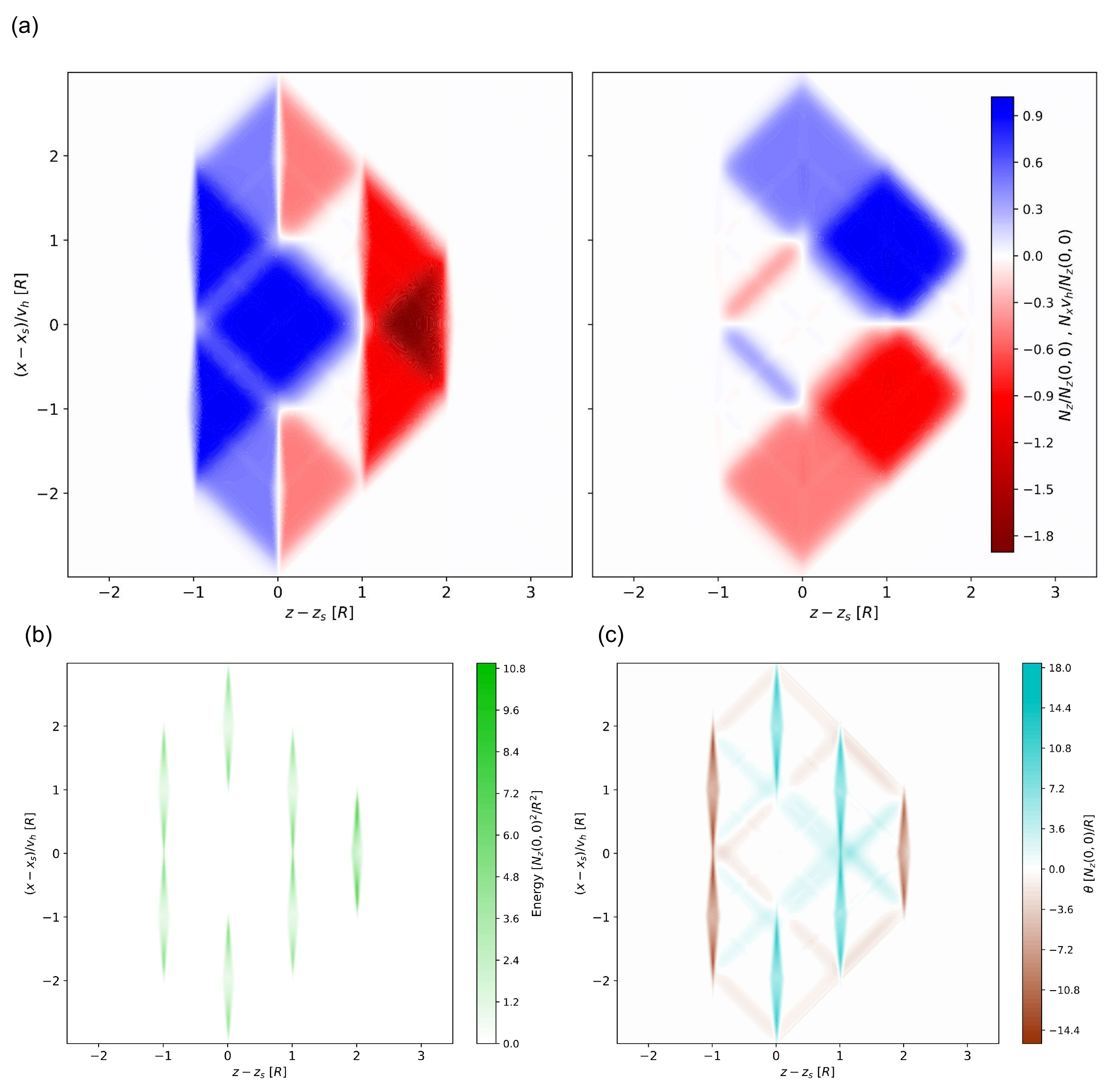}
\caption{(a) Projection of the shift vector components $N_z$ (left) and $N_x$ (right) along $(x,0,z)$. Propagation direction of the soliton is from left to right along the $z$-axis. The multi-compartment structure is a distinct departure from the single top-hat soliton found in Alcubierre 1994~\cite{Alcubierre1994} and Nat\'ario 2002~\cite{Natario2002}. Total integrated shift in each direction is 0. Note that the shift vector components are normalized with respect to the value of $N_z$ a the co-moving origin. 
(b) Projection of the local energy density. The energy density is dominated by those regions containing hyperbolic source $\rho$, but also extends weakly to the boundaries of the wavefronts. The energy density is everywhere positive for Eulerian observers.
(c) Projection of the local volume expansion factor $\theta$. Positive and negative expansion factor are largely associated with negative and positive hyperbolic sources respectively. Non-zero expansion factor also exist in the spaces in-between hyperbolic sources along the hyperbolic wavefronts. Total integrated expansion factor is 0.
These plots are taken from Lentz 2021~\cite{Lentz2020}.}
\label{fig:Ns}
\end{center}
\end{figure}

The shift vector of the positive-energy soliton created in Lentz 2021~\cite{Lentz2020} is given in Fig.~\ref{fig:Ns}. The soliton moves along the positive z axis at a speed set by the value of the shift vector at the origin of the co-moving coordinate coordinates in Fig.~\ref{fig:Ns}, which may be given arbitrary positive value. The transport logistics of the solitons are then similar to that of the Alcubierre solution. The solitons are constructed to contain a central region with minimal tidal forces, where proper time coincides with asymptotic coordinate time, and any Eulerian observer -- which in this case is free falling and whose velocity matches the shift vector -- within the central region would remain stationary with respect to the soliton. This is the region where a spacecraft would be placed.

The total energy requirements of the positive-energy solitons closely follow that of Pfenning \& Ford 1997~\cite{Pfenning1997} as applied to the Alcubierre solution
\begin{equation}
    E_{tot} = \int E \sqrt{-g} d^3x.
\end{equation}
For solitons where the radial extent of the central region $R$ is much larger than the thickness of the energy-density laden boundary shell $w$ ($w \ll R$), the energy is estimated to be
\begin{equation}
    E_{tot} \sim  C v_s^2 \frac{R^2}{w}
\end{equation}
where $C$ is a form factor typically of order unity. The required energy for a positive-energy soliton with central region mean radius $R = 100$~m and average source thickness along the z-axis $w = 1$~m approaches a mass equivalent of $E_{tot} \sim (\text{few}) \times 10^{-1} M_{\odot} v_s^2$, which is of the same magnitude as the estimate of an Alcubierre solution of the same dimensions.

\section{Addressing the Literature}

The findings of Lentz 2021~\cite{Lentz2020} run against the common wisdom of the warp drive literature to that point and the proofs set forth in Olum 1999~\cite{Olum1998} and Lobo \& Crawford 2003~\cite{Lobo2003} stating that any superluminal spacetime must violate the WEC via violations of the null energy condition (NEC). The proofs are both based on an analysis of the Raychaudhuri equation for null geodesics, confined to spacetimes with only a single fastest (superluminal) causal path between two space-like 2-surfaces. The pre-conditions of these proofs are very local in nature and appear analogous to collapsing the interior of a warp drive soliton to a point in order to produce a single fastest causal path. The solitons of the early literature, such as Alcubierre 1994~\cite{Alcubierre1994} and Nat\'ario 2002~\cite{Natario2002}, have simple structures and can survive this limit. The example positive energy warp drive of Lentz 2021~\cite{Lentz2020} cannot undergo this limit without being destroyed. This drive therefore does not meet the pre-conditions of the proofs and exists outside their scope, implying that the proofs are not applicable.

Several warp drive papers have addressed the findings of Lentz 2021~\cite{Lentz2020} since an early manuscript of it was made publicly available \cite{Fell2021,Bobrick2021,Santiago2021a,Santiago2021a,SantosPereira2021b}. Of particular note, the papers of \cite{Santiago2021a, Santiago2021b} have made several assertions claiming that the solution of Lentz 2021~\cite{Lentz2020} cannot satisfy the WEC. Follow-up correspondence with the authors as well as discussion captured at the recent Marcel Grossmann meeting have demonstrated that these papers did not adequately analyze the contents of Lentz 2021~\cite{Lentz2020}. To summarize the discussions, Santiago et al.~\cite{Santiago2021a,Santiago2021b} argue that the Eulerian energy density of a soliton in a Nat\'ario class spacetime can be written as the sum of a total divergence and a negative definite term
\begin{equation}
    E = \frac{1}{16 \pi} \left( \partial_i \left(N_i \partial_j N_j - N_j \partial_j N_i \right) - \frac{1}{2} \omega_i \omega_i \right),
\end{equation}
where $\omega_i = \epsilon_{ijk} \partial_k N_j$ is the shift vector vorticity. The divergence term is then argued to produce zero net energy if the warp drive is finite in size due to an application of the divergence theorem on the hypersurface where the integral's volume boundary is extended towards infinity where the divergence kernel quickly vanishes, implying that the total Eulerian energy of a Nat\'ario spacetime is non-positive. This argument does not hold in the case of Lentz 2021~\cite{Lentz2020} as the Eulerian energy density of the example positive energy soliton is smooth save for the boundaries $x=0$ and $y=0$ where stress-energy sources are only continuous, while a requirement of the divergence theorem is that the total divergence be at least first order smooth everywhere. The integral volume boundary therefore cannot be separated from the soliton and instead must be applied in a patchwork~\cite{Dray1994}, with some boundaries running adjacent to the 2-surfaces $x=0$ and $y=0$, where the divergence kernel is non-vanishing. Expansion beyond Nat\'ario class of spacetimes may smooth the geometry and sources of positive energy warp drives further.

\section{Further Challenges and Future Prospects}

There are still numerous challenges between the current state of physical warp drive research and a functioning prototype. I list here several of the more near-term challenges and give my perspective as to how research in these areas may be approached.

The most glaring challenge is the astronomical energy cost of even a modest warp drive, currently measured in solar masses where kilograms is closer to the threshold of human technology. Extreme energy savings is going to be necessary -- tens of orders of magnitude -- to bring the energy required for a warp drive down to a level that can be tested in a laboratory setting let alone be considered a viable transportation technology.

There exist numerous techniques for reducing the energy requirements of the Alcubierre solution, several of which have been very successful in reducing the (magnitude) energy requirements of the system in excess of thirty orders of magnitude \citep{VanDenBroeck1999,Loup2001,Krasnikov2003,Obousy2008,White2013}. Unfortunately, each one of these methods in their presented forms require negative energies themselves. One possible approach to uncovering significant energy savings is to modify one of these existing techniques to obey the WEC.

If the required energy can be sufficiently reduced, the next hurdle to approach is modeling the full life cycle of a physical warp drive (creation, acceleration, inertial motion, deceleration, and diffusion). Every previous publication in the field of warp drives has either assumed inertial motion (constant velocity) or has produced an accelerating/decelerating drive that violates the law of covariant conservation of stress-energy-momentum
\begin{equation}
    \mathbf{\nabla} \cdot \mathbf{T} = \mathbf{0},
\end{equation}
that accompanies the Einstein equation. Deriving mechanisms for creation and acceleration is crucial to any experimental test.

The last hurdle I will mention is the full characterization of the sourcing fields, whether it be a plasma or other state of matter and energy. As stated by Matt Visser in the Q\&A of my talk, the specification of the drive geometry only is an incomplete description of the full solution. Stress-energy sources must be specified to close the system. In the hypothetical plasma of Lentz 2021~\cite{Lentz2020}, the stress-energy governing equations include the Maxwell equations for the electric and magnetic fields, the equations of motion for each species of matter, and various constituent equations governing the state of the Einstein-Maxwell-matter system. The total system is expected to be far too complex to provide analytical solutions, requiring numerical simulation as the primary means to specify each field of a soliton at any point in its life cycle.

\bibliographystyle{ws-procs961x669}
\bibliography{ws-pro-sample}

\end{document}